\documentclass[aps,prl,reprint,superscriptaddress]{revtex4-1}

\usepackage{amsmath}
\usepackage{amssymb}
\usepackage{graphicx}
\usepackage{array}
\usepackage{subeqnarray}
\usepackage{blindtext}
\usepackage{float}

\begin{document}


\title{Symmetry Plays a Key Role in the Erasing of Patterned Surface Features}



\medskip

\author{Michael Benzaquen}\thanks{These authors contributed equally to this work.}
\affiliation{UMR CNRS Gulliver 7083, ESPCI ParisTech, PSL Research University, Paris, France}
\author{Mark Ilton}\thanks{These authors contributed equally to this work.}
\affiliation{Department of Physics \& Astronomy, McMaster University, Hamilton, Ontario, Canada}
\author{Michael V. Massa}
\affiliation{Department of Physics \& Astronomy, McMaster University, Hamilton, Ontario, Canada}
\author{Thomas Salez}
\affiliation{UMR CNRS Gulliver 7083, ESPCI ParisTech, PSL Research University, Paris, France}
\author{Paul Fowler}
\affiliation{Department of Physics \& Astronomy, McMaster University, Hamilton, Ontario, Canada}
\author{Elie Rapha\"{e}l}
\affiliation{UMR CNRS Gulliver 7083, ESPCI ParisTech, PSL Research University, Paris, France}
\author{Kari Dalnoki-Veress}\email{dalnoki@mcmaster.ca}
\affiliation{Department of Physics \& Astronomy, McMaster University, Hamilton, Ontario, Canada}
\affiliation{UMR CNRS Gulliver 7083, ESPCI ParisTech, PSL Research University, Paris, France}



\medskip


\begin{abstract}
\medskip

We report on how the relaxation of patterns prepared on a thin film can be controlled by manipulating the symmetry of the initial shape. The validity of a lubrication theory for the capillary-driven relaxation of surface profiles is verified by atomic force microscopy measurements, performed on films that were  patterned using focused laser spike annealing. In particular, we observe that the shape of the surface profile at late times is entirely determined by the initial symmetry of the perturbation, in agreement with the theory. Moreover, in this regime the perturbation amplitude relaxes as a power-law in time, with an exponent that is also related to the initial symmetry. The results have relevance in the dynamical control of topographic perturbations for nanolithography and high density memory storage.

\end{abstract}


\maketitle


\subsection{Introduction}

Thin polymer films are of general interest, being both industrially relevant and readily amenable to experiment~\cite{Tsui2008}. Used in diverse applications such as data storage, lubricant coatings, electronic devices, and wire arrays, polymer films can be easily tuned in both their wetting properties as well as their dynamics. An area of especially active research involves the use of thin polymer films for nanoscale pattern templating. Block copolymer lithography~\cite{Nunns2013,Boyd2013,Tseng2010,Marencic2010,Hamley2009}, for instance, has been used to shape samples on sub-10~nm length-scales~\cite{Park2008,Son2011,Bates2012} by taking advantage of the self-assembly of amphiphilic polymer molecules. This self-assembly can be further controlled by topographic perturbations, for example those created using graphoepitaxy or grayscale lithography, on larger mesoscopic length-scales~\cite{Cheng2004,Cheng2006a,Bita2008}. Topographic perturbations can also be used to directly pattern homogeneous thin films{,} as is the case in nanoimprint lithography~\cite{Chou1995,Chou1996a,Austin2002,Guo2004,Guo2007}, {and is applicable} as a data storage technique with dense memory capabilities~\cite{Vettiger2002,Pozidis2006}, in self-cleaning surfaces~\cite{Bixler2013}, and organic optoelectronics~\cite{Kim2012,Bay2013}. The relaxation of thin film perturbations has been used to study glassy polymer dynamics~\cite{Fakhraai2008,Yang2010,Chai2014}, film viscosity~\cite{Leveder2008,Leveder2011,Rognin2011,McGraw2012}, and viscoelastic properties~\cite{Rognin2012,Benzaquen2014,Rognin2014}. In essence, topographic perturbations can be used not only in patterning films for applied technologies, but as a way to study material properties on small length-scales that are inaccessible with bulk measurement techniques.\smallskip

Perturbations can be created atop a polymer film, on a mesoscopic length-scale in a variety of ways. Unfavourable wetting properties~\cite{Srolovitz1986,Wyart1990,Seemann2001a,Reiter2005a,Chen2012}, electro-hydrodynamic instability~\cite{Schaffer2000,Morariu2003,Voicu2006}, Marangoni flow~\cite{BinKim2014,Katzenstein2014,Arshad2014,Katzenstein2012}, and thermocapillary forces~\cite{Brochard1989,Kataoka1999,Valentino2005,Dietzel2009,Singer2013} can all drive a flat film away from a uniform film thickness. The film viscosity $\eta$, surface tension $\gamma$, and unperturbed film thickness $h_0$, are three parameters that influence the effective mobility of a film, which affects the relaxation of an applied surface perturbation. A time-scale $t_0 = 3\eta h_0 / \gamma$ can be used to characterize the relaxation of a viscous film~\cite{Salez2012a}. By increasing the temperature or placing the film in solvent vapour, the effective mobility of the film can be increased{,} causing a faster relaxation of topographic perturbations. Finally, geometry appears to play a key role as well, since a long and straight trench\cite{Baumchen2013} relaxes with a different power-law in time than a cylindrical mound\cite{Backholm2014}. \smallskip

In this article, we rationalize a new method to control the surface relaxation rate of a thin film, based on the geometrical properties of its initial pattern. First, we present a linear theory of the capillary relaxation of surface profiles. Then, the validity of the asymptotic series expansion of the general solution is experimentally tested using focused laser spike annealing and atomic force microscopy. In agreement with theory, we find the shape and relaxation rate of surface features to strongly depend on their initial symmetry. More specifically, within the configurations studied here, we observe that quickly erasable features can be created by patterning an initial perturbation with a high degree of spatial symmetry.

\subsection{Theoretical Results}
For an annealed film with a vertical thickness profile described by $h(\boldsymbol{r},t) = h_0 + d(\boldsymbol{r},t)$, the surface displacement $d(\boldsymbol{r},t)$ at a given horizontal position $\boldsymbol{r}$ decays in time $t$ due to capillary forces, and the final equilibrium state is a flat film with uniform thickness $h_0$.

\subsubsection{2D Case}
When the system is bidimensional, namely invariant along one spatial direction, the perturbation is a function of one spatial direction $x$ only, and $\boldsymbol{r}$ is replaced by $x$. In such a case, one can show within a lubrication model (see Theoretical Methods) that the perturbation is given by the asymptotic series expansion:
\begin{eqnarray}\label{2D_expansion}
\frac{d(x,t)}{h_0}=\underbrace{\frac{\mathcal M_0F_0(u)}{(t/t_0)^{1/4}}}_{\substack{\text{non-zero}\\\text{volume}}}-\underbrace{\frac{\mathcal M_1 F_1(u)}{(t/t_0)^{1/2}}}_{\substack{\text{zero-volume}\\\text{asymmetric}}}+\underbrace{\frac12\frac{\mathcal M_2 F_2(u)}{(t/t_0)^{3/4}}}_{\substack{\text{zero-volume}\\\text{symmetric}}}-  \dots \ \ \
\end{eqnarray}
where $u=(x/h_0)/(t/t_0)^{1/4}$ is a dimensionless variable. Each term in the above infinite expansion has a dimensionless attractor function $F_i(u)$, and two prefactors: the moment $\mathcal M_i/(i!)$ and the temporal dependence $(t_0/t)^{(i+1)/4}$. The prefactors are functions of the initial state of the perturbation and the characteristic time-scale $t_0$. In the first term, $\mathcal M_0$ is proportional to the amount of excess volume the perturbation adds, or equivalently the 0\textsuperscript{th} moment of the initial profile ($\mathcal M_0 \propto \int \text dx\, d(x,0)$). For that reason, this term is labelled as `non-zero volume'. In the second term, $\mathcal M_1$ is non-zero when the profile is asymmetric, and it is proportional to the 1\textsuperscript{st} moment of the initial profile ($\mathcal M_1 \propto\int  \text dx\, x\, d(x,0)$). Because, at long times, this term becomes the leading order term when $\mathcal M_0 = 0$ and $\mathcal M_1\neq 0$, it is termed `zero-volume asymmetric'. Similarly, the third term is proportional to the 2\textsuperscript{nd} moment of the initial distribution ($\mathcal M_2 \propto\int  \text dx \, x^2\, d(x,0)$). This term becomes dominant when the initial distribution has no excess volume and is perfectly symmetric, namely $\mathcal M_0=\mathcal M_1=0$ and $\mathcal M_2\neq 0$, and is thus labelled `zero-volume symmetric'. The attractor functions {$F_i(u)=F_0^{(i)}(u)$ are the $i$-th derivatives of $F_0(u)$ and encode the rescaled shape of the spatial profile of the perturbation. Examples of attractor functions are shown in Figure~1(b,d). Because of the different power-laws in time for each term, after an initial transient regime the overall relaxation is dominated by the first term with a non-zero prefactor in Eq.~\eqref{2D_expansion}, regardless of the exact shape of the initial surface feature. In that sense, the attractor functions are referred to as \textit{universal} attractors. If volume is added to the reference flat film by the initial perturbation, $\mathcal M_0 \neq 0$, the profile $d(x,t)$ will converge to the function $F_0(u)$ in finite time \cite{Benzaquen2014}. If no volume is added by the initial perturbation, then the profile will converge to the first term with a non-zero prefactor in Eq.~\eqref{2D_expansion}.

\subsubsection{3D Case}
In the case where the surface displacement is a function of two spatial dimensions in the plane, $\boldsymbol{r}=(x,y)=(r,\psi)$, an angular average on $\psi$ around the center \footnote{In the general case, although the center of the surface perturbation is not uniquely defined, the results are independent of the choice of the center. 
For most real experimental features, the position of the center {can be chosen} quite naturally.}  of 

\begin{widetext}

\begin{figure}[t!]
\includegraphics[width=125mm]{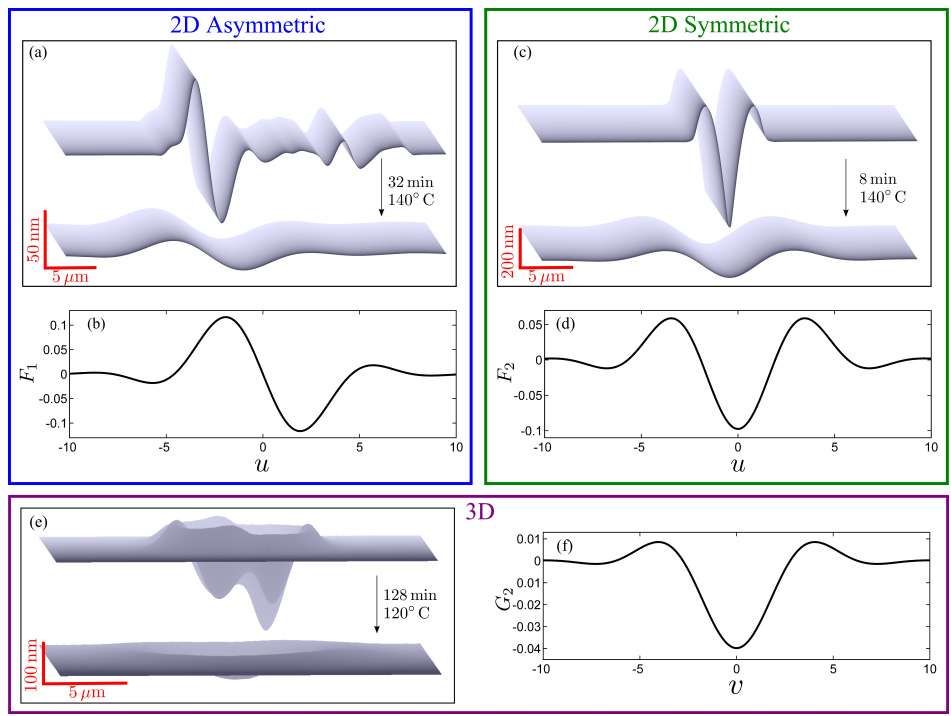}
\caption{(a,c,e) Experimentally measured AFM profiles of three different zero-volume surface perturbations atop thin polystyrene films, initially and after annealing (see Experimental Methods). (b,d,f) Corresponding attractor functions which appear in Eqs.~\eqref{2D_expansion} and \eqref{3D_expansion} (see Theoretical Methods).\label{FigAFM}}
\end{figure}

\end{widetext}

 \noindent the perturbation can be taken and the averaged profile, $\langle d(r,{\psi},t) \rangle_{\psi}$, can be written (see Theoretical Methods)  as the following asymptotic series expansion:
\begin{eqnarray}\label{3D_expansion}
\frac{\langle d(r,{\psi},t) \rangle_{\psi}}{h_0}&=&\underbrace{\frac{\mathcal N_0G_0(v)}{(t/t_0)^{1/2}}}_{\substack{\text{non-zero}\\\text{volume}}}+\underbrace{\frac12\frac{\mathcal N_2G_2(v)}{t/t_0}}_\text{zero-volume}+ \dots \ .
\end{eqnarray}
Here, the dimensionless variable reads $v=(r/h_0)/(t/t_0)^{1/4}$, where $r$ is the radial distance from the center. Each term in the series has a similar structure to that of the 2D case. There are moments $\mathcal N_i$ that depend on the symmetry of the initial perturbation, attractor functions $G_i(u)$ that encode the rescaled shape of the spatial profile, and  power-laws in time which have larger respective exponents than the 2D case. Higher order terms in the series are zero-volume terms that depend on higher order moments of the initial perturbation.

Previous studies have focused on {non-zero volume perturbations and} the convergence to 0\textsuperscript{th} order terms in the 2D ~\cite{Benzaquen2013,Baumchen2013,Benzaquen2014} and 3D cases~\cite{Backholm2014}. In particular, special attention was dedicated to the convergence time\cite{Benzaquen2014}, a first crucial quantity for practical purposes as it is the time-scale after which a surface feature has undergone significant relaxation. In the present article, we instead study zero-volume perturbations which are of technological  significance, since patterns designed by inducing flow in the material (\textit{e.g.} through wetting properties, electro-hydrodynamic instabilities, or thermocapillary forces) show no volume change from the initially flat film. In that case, the first terms in Eqs.~\eqref{2D_expansion} and \eqref{3D_expansion} vanish -- $\mathcal M_0 = 0$ in the 2D case, and $\mathcal N_0 = 0$ in the 3D case -- and the relaxation at late times is therefore dominated by the next, lowest, non-zero moment. In contrast with previous investigations on the convergence time\cite{Benzaquen2014}, we here concentrate our efforts on the second crucial quantity for practical purposes: the temporal exponent of the relaxation. For zero-volume perturbations with  similar convergence times, we show that the higher the symmetry, the larger the exponent \textit{i.e.} the faster the erasing.

\subsection{Experimental Results}
In order to test the role symmetry plays in surface relaxation, three types of zero-volume perturbations were made (see Figure~\ref{FigAFM}) using focused laser spike annealing~\cite{Hudson2004,Parete2008,Singer2013} on thin polystyrene films: i) a 2D asymmetric feature which maximizes the second term in Eq.~\eqref{2D_expansion}, as shown in Figure~1(a,b); ii) a 2D symmetric feature dominated by the third term in Eq.~\eqref{2D_expansion}, as shown in  Figure~1(c,d); and iii) a 3D feature with no apparent symmetry such that only the first term in Eq.~\eqref{3D_expansion} is zero, as shown in Figure~1(e,f). The 2D asymmetric feature was created with large extrema, with a peak-to-peak height of 114 nm (see Experimental Methods). The 2D symmetric feature had an initial amplitude of 361 nm. Finally, the 3D feature was created by making 4 different depressions of varying depths, resulting in a deepest feature with an amplitude of 188 nm, and with three smaller features nearby.

%

Each sample was annealed above the glass transition temperature to probe the surface relaxation. After a certain annealing time, the sample was quenched to room temperature, its height profile measured with atomic force microscopy (AFM), and then it was placed back on the hot stage in order to repeat the annealing-quenching-measure sequence. The 2D features were annealed at $140^{\circ}\,\textrm{C}$. Because the 3D dynamics is faster than the 2D  case, as stated above, the 3D feature was annealed at $120^{\circ}\,\textrm{C}$ to slow down its relaxation and ensure that the annealing times were much longer ($\geq 1$ min) than the time it took to quench the sample ($<10$ s).
\begin{figure}[t!]
\includegraphics[width=87mm]{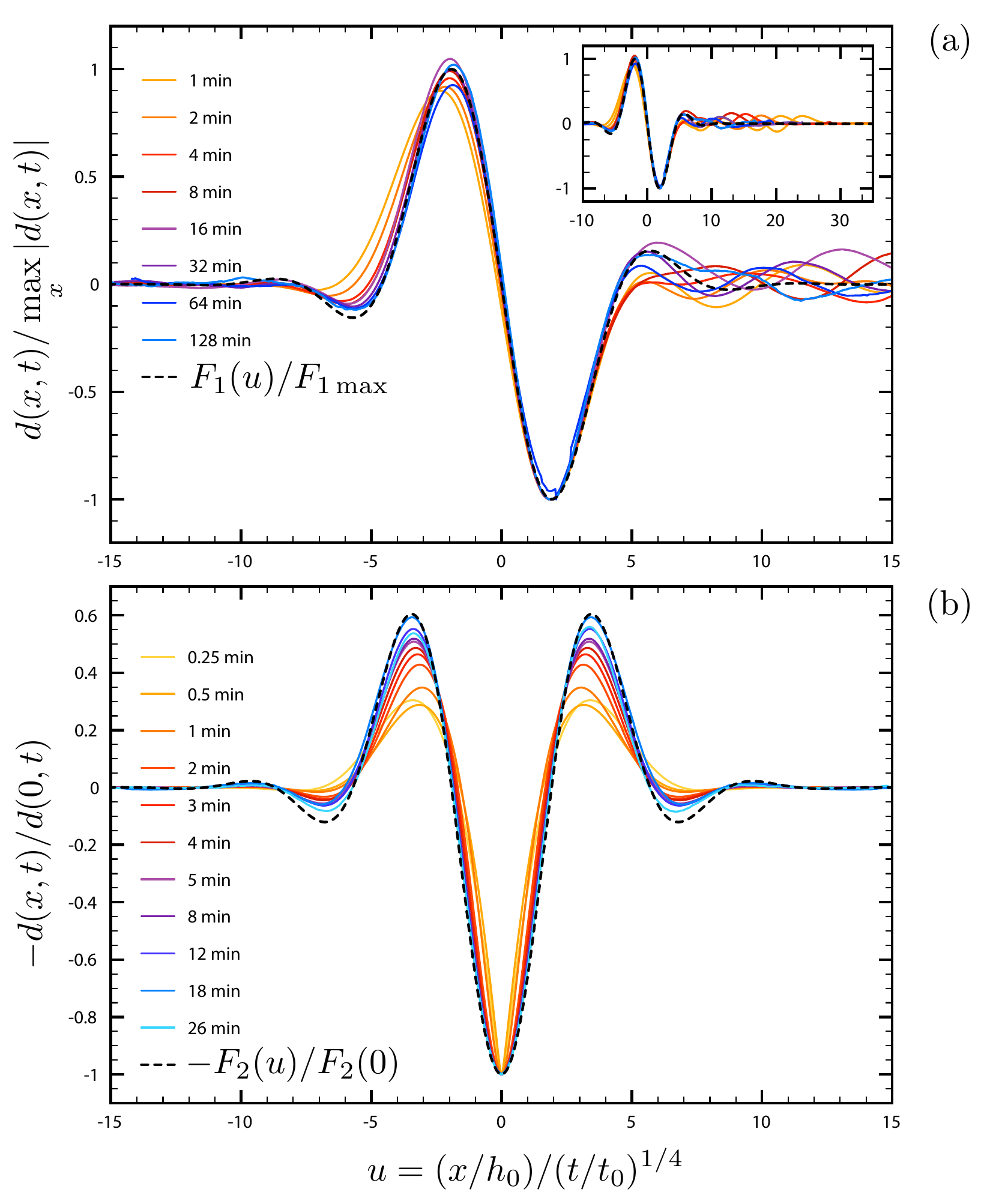}
\caption{Normalized profiles of 2D features for both the (a) asymmetric and (b) symmetric initial perturbations, as a function of the rescaled horizontal position $u=(x/x_0)/(t/t_0)^{1/4}$, for different annealing times $t$. The black dashed lines correspond to the normalized attractor functions in 2D (see Eq.~\eqref{2D_expansion},  Figs.~\ref{FigAFM}(b) and (d), and Theoretical Methods). The inset shows that the oscillations on the right hand side have a finite spatial extent.}
\end{figure}

\subsubsection{Convergence of the profiles to the attractor functions}
To explicitly test the convergence of the 2D surface profiles to the corresponding attractor functions $F_i(u)$, the normalized profiles are plotted in Figure 2 as a function of the rescaled position $u$ for several times $t$. Figure~2(a) shows the normalized relaxation profile of the 2D asymmetric feature from Figure~1(a), that is with $\mathcal M_0 = 0$ and $\mathcal M_1 \neq 0$. The profiles collapse onto the normalized attractor $F_1/F_{1\max}$, which corresponds to the lowest-order non-zero term from Eq.~(\ref{2D_expansion}). Similarly, Figure~2(b) shows the normalized relaxation profile of the 2D symmetric feature from Figure~1(c), that is with $\mathcal M_0= \mathcal M_1=0$ and $\mathcal M_2 \neq 0$. In this case, the data collapses to the normalized attractor $F_2/F_2(0)$.

For the 3D feature shown in Figure~1(e), that is with $\mathcal N_0=0$, the normalized profiles are shown from above in Figure~3(a) along with the normalized attractor $G_2(v)$. The feature starts off with a low degree of symmetry, evolving towards a roughly axisymmetric depression. The radially averaged profiles are shown in Figure~3(b), with $v=0$ taken to be the deepest point of the surface perturbation. As predicted by Eq.~\eqref{3D_expansion}, there is a collapse of the profiles to $G_2(v)$ at late times.

\subsubsection{Influence of symmetry on the relaxation dynamics}
So far, we have shown that depending on the initial symmetry of a surface perturbation, Eq.~\eqref{2D_expansion} or Eq.~\eqref{3D_expansion} describes well the \textit{shape} of the relaxing profile. Now, we focus on the temporal evolution of the \textit{amplitude} of such perturbations. We show that the initial symmetry plays a key role in the relaxation rate. According to Eqs.~\eqref{2D_expansion} and \eqref{3D_expansion}, the maximum amplitude $d_{\textrm{max}}=\textrm{max}(\lvert d\rvert)$ of the perturbation should scale as a power-law in time $t$ for sufficiently long times. The power-law exponent should  depend on which order/term controls the relaxation.

\begin{widetext}

\begin{figure}
\centering
\includegraphics[width=115mm]{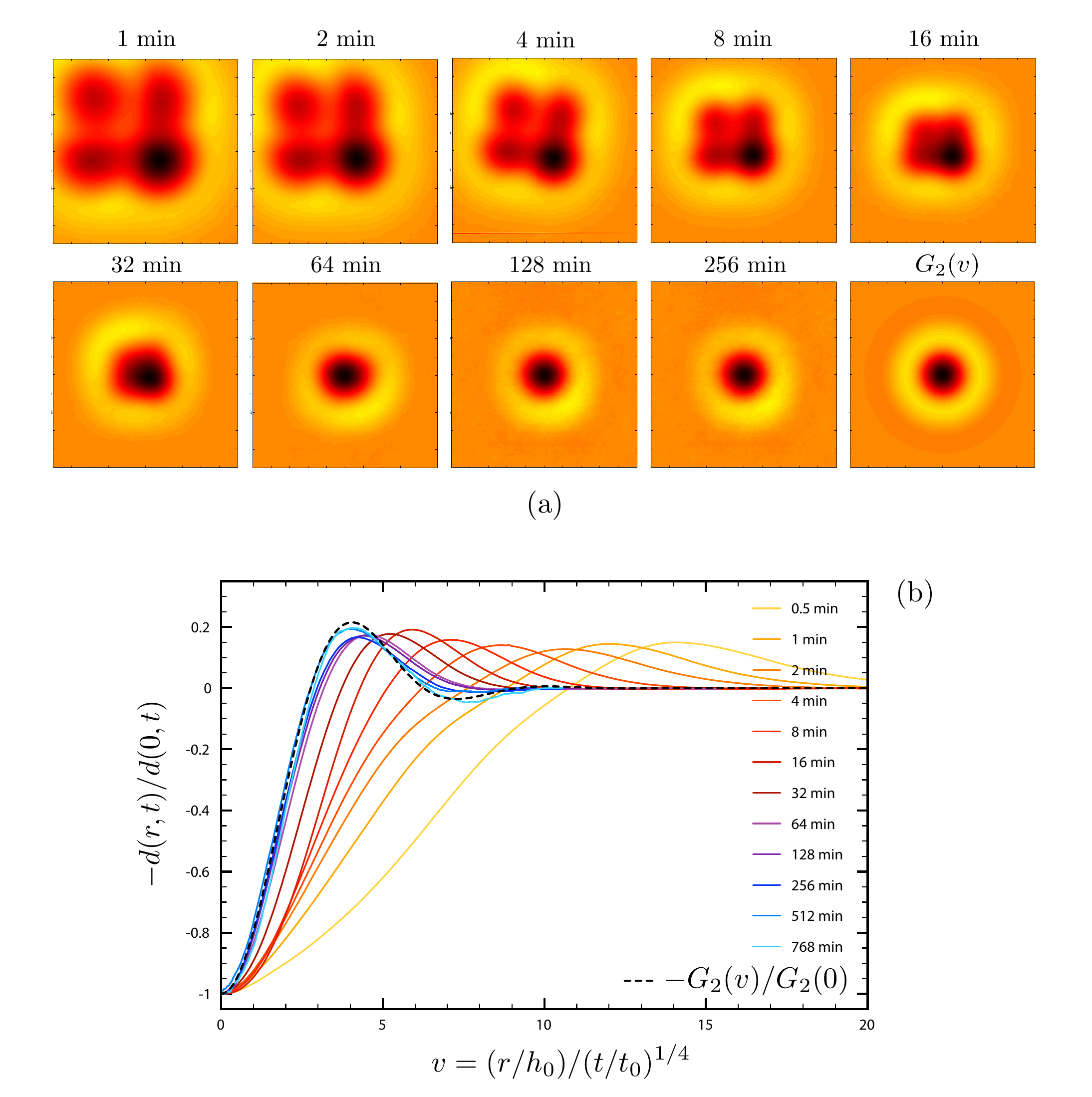}
\caption{(a) AFM images showing the temporal evolution of the 3D perturbation viewed from above (see Figure~\ref{FigAFM}(e)), shown in normalized horizontal units,  $(x/h_0)/(t/t_0)^{1/4}$ and $(y/h_0)/(t/t_0)^{1/4}$, and color-scaled by the amplitude of the feature: the darker the deeper. The corresponding 3D attractor (see Eq.~\eqref{3D_expansion}, Figure~\ref{FigAFM}(f), and Theoretical Methods) is shown in the last panel. (b) Angularly averaged profiles of the normalized AFM images above plotted as a function of the rescaled radius $v$ and compared to the attractor.}
\end{figure}

\end{widetext}

\noindent  For example, for the 2D asymmetric feature, the $F_1(u)$ term is dominant which means Eq.~\eqref{2D_expansion} predicts $d_{\textrm{max}} \sim t^{-1/2}$ at late times. {Since $\eta$ and $h_0$ are obvious factors controlling the dynamics of the film, they have been scaled out using a normalized time.  Here, we use the convergence time $t_\textrm{c}$, which was previously defined\cite{Benzaquen2014} as the time when the asymptotic power-law behavior for the amplitude equals the initial amplitude of the perturbation. Convergence times can been computed theoretically for the three present configurations using a similar definition (see Theoretical Methods), and the experimentally determined convergence times are given in Fig.~\ref{dmaxFig}.} The normalized amplitude $d_{\textrm{max}}(t)/d_{\textrm{max}}(0)$ is thus plotted against the normalized time $t/t_\textrm{c}$ in Fig.~\ref{dmaxFig}. The late-time relaxation of each of the three data sets agrees very well with the theoretical power-law predictions from Eqs.~\eqref{2D_expansion} and \eqref{3D_expansion}. The 3D feature has the steepest relaxation, followed by the 2D symmetric feature, and finally the 2D asymmetric feature. Note that, in the current work, we observe $t_\textrm{c}/t_0$ to be smaller for patterns described by a steeper power-law (Fig.~\ref{dmaxFig}, legend), but in general this is not the case for an arbitrary initial profile shape. There are in fact two independent key control parameters: the convergence time and the time exponent.\\

To sum up, the relaxation of zero-volume perturbations on a flat thin film agrees very well with linear lubrication theory. Perturbations with lower symmetry and dimensionality were observed to have the slowest evolution, while perturbations with higher symmetry and dimensionality relaxed more quickly. This has clear implications for the use of topographic perturbations in an applied context. At fixed temperature (or viscosity) and film thickness, if the goal is to create a liquid perturbation stabilized against flow, one should aim to have the lowest possible symmetry and dimensionality. On the other hand, if a quickly erasable perturbation is desirable, the use of higher symmetry and higher dimensionality (3D, rather than 2D feature) would give a faster relaxation. Reliably creating 3D perturbations with a large fourth-order moment, but with zero lower-order moments, is technically challenging but would allow for fast erasing processes. Such strategies, and the fine design of properly shaped nanoindentors and masks, would speed up -- and thus improve --  nanomechanical memory storage~\cite{Vettiger2002,Pozidis2006}, at given temperature and film thickness.\\

In conclusion, we have used focused laser spike annealing to create zero-volume surface perturbations in thin polystyrene films. The relaxation of the initial profiles was measured as a function of time, as the film was driven by surface tension towards the equilibrium state of a flat film. We have shown that the surface relaxation agrees very well with a linear lubrication theory. In particular, the surface profiles collapse to predicted attractor functions in both 2D and 3D. The amplitudes of the features follow a power-law relaxation in time, the exponent of which is determined by the lowest moment of the initial profile. We have discovered a new strategy for tuning the stability and relaxation capabilities of patterned features at the nanoscale. The dimensionality and initial symmetry play a crucial role in the relaxation time-scale of a thin film perturbation. 
A stable liquid feature is created by adding -- or removing -- material on an initially flat film and by choosing a 2D initial profile shape with a large convergence time. A quickly erasable feature needs to be patterned in 3D with a short convergence time, and a high degree of symmetry.

\begin{figure}[t!]
\includegraphics[width=87mm]{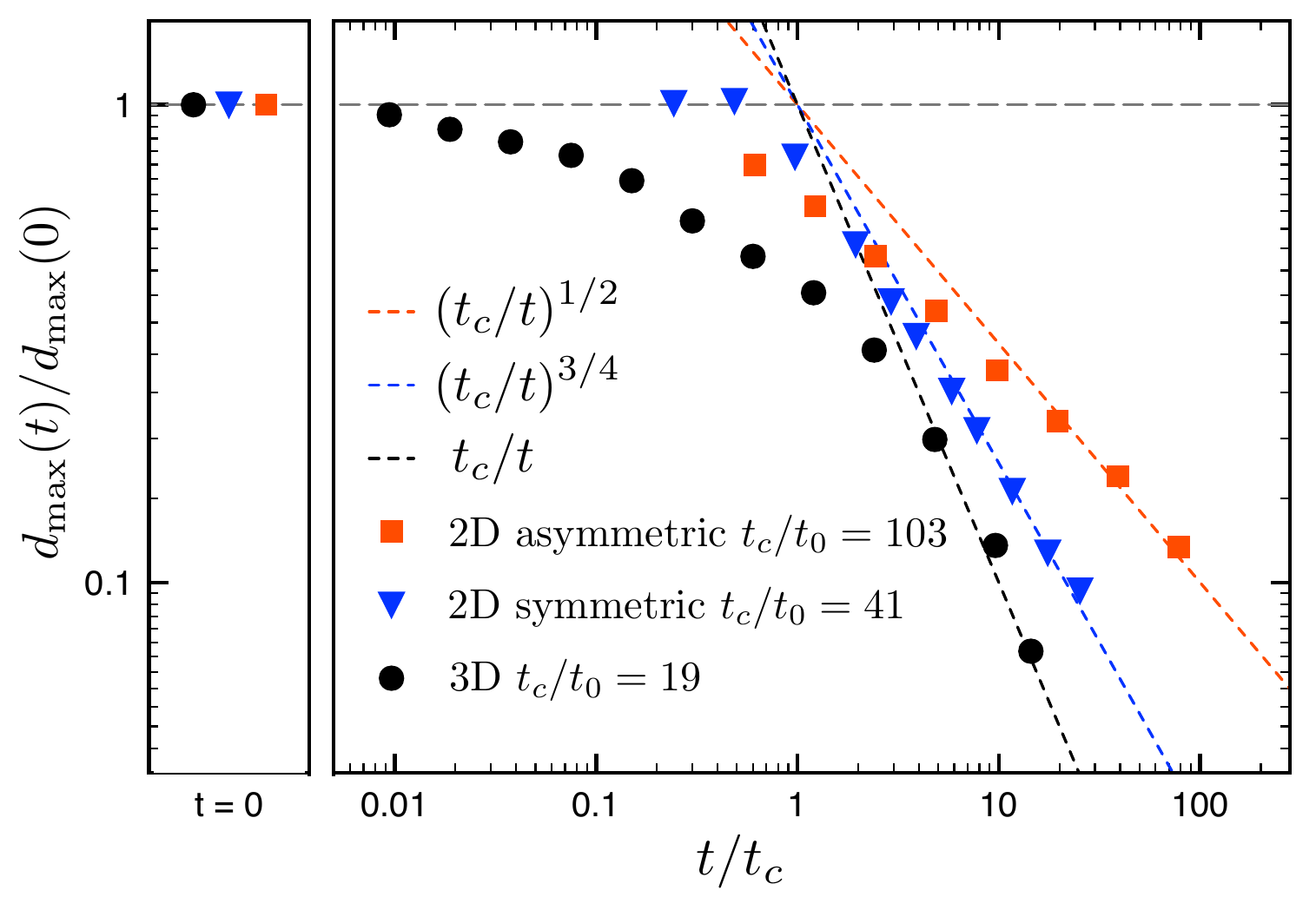}
\caption{Double logarithmic plot of the normalized amplitude $d_{\textrm{max}}/d_{\textrm{max}}(0)$ of the three different zero-volume surface perturbations, as a function of time $t$ normalized by the convergence time $t_\textrm{c}$ (solid points). The left panel shows the normalized amplitudes at $t=0$. The dashed lines are the corresponding power-laws predicted by Eqs.~\eqref{2D_expansion} and \eqref{3D_expansion}.} \label{dmaxFig}
\end{figure}

\subsection{Acknowledgements}
Financial Support for this work was provided in part by NSERC (Canada). The authors thank Mark Ediger for an interesting discussion on this topic.

\subsection{Experimental Methods}
Thin polystyrene films, with molecular weight $M_w = 31.8$ kg/mol and polydispersity index 1.06 (Polymer Source), were spincast from dilute toluene solution (Fisher Scientific, Optima grade) onto silicon wafers (University Wafer). The films were pre-annealed for 2 hours at $150^{\circ}\,\textrm{C}$ on a hot stage (Linkham Scientific Instruments) to relax the polymer chains. Zero-volume surface perturbations were created using a home-built focused laser spike annealing setup similar to that described previously~\cite{Hudson2004,Parete2008,Singer2013}. Briefly, a focused laser (Coherent, Verdi V2, 532 nm) is rastered across the thin polymer film at room temperature. The silicon substrate absorbs some of the laser energy which creates a large temperature gradient. This locally heats the polymer film above its glass transition ($\sim 100\ ^\circ$C). Since the surface tension decreases with increasing temperature, there is a surface tension gradient that drives flow away from the region of higher temperature, thus creating a perturbation in the film without any resulting change in volume. Both 2D and 3D depressions can be created using this method. A 2D feature is created by holding the laser at a constant power and moving it over the surface at constant speed in a long (200 $\mu$m) straight motion, creating a nearly-uniform height profile in the direction $y$ parallel to the laser motion. The 2D asymmetric feature was created using multiple passes of the laser with progressively lower power and a small horizontal shift along $x$ between passes. As a line is rastered with the laser, material gets pushed to either side of the line. Because there are multiple passes of the laser, there is an imperfect clearance of the material resulting in small oscillations to the right ($x>0$). These are all less than $\sim20\%$ of the main feature height (see inset of Figure~2(a)), and become less important after annealing. On the other hand, the 2D symmetric feature was created using a single pass of the laser, and no oscillation was present due to the single pass of the laser. Finally, a 3D depression is created by opening the laser shutter at a fixed point for a brief amount of time ($\approx 1$ s). Atomic force microscopy (AFM, Veeco, Caliber) was used to measure the surface profiles of the films and was performed after a quench at room temperature.

\subsection{Theoretical Methods}
The theory is based on the lubrication approximation and the thin film equation \cite{Blossey2012}:
\begin{eqnarray}
\partial_th+\frac{\gamma}{3\eta}\,  \boldsymbol \nabla \cdot  \left( h^3\,\boldsymbol \nabla \Delta h \right)&=&0\ ,
\label{TFEd}
\end{eqnarray}
which describes the capillary-driven relaxation of a thin supported film with vertical thickness profile $h(\boldsymbol{r},t)$, along horizontal space $\boldsymbol r$ and time $t$.
Equation \eqref{TFEd} can be nondimensionalized through $h=h_0+d(\boldsymbol r,t)=H\, h_0$, $\boldsymbol r= \boldsymbol R\, h_0$ and {$t=T\,t_0$, where $t_0={3\eta h_0}/{\gamma}$, and}
where $h_0$ is the reference height at infinity. Equation~\eqref{TFEd} is highly nonlinear and, as of today, hasn't been solved analytically. When the surface of the film is only slightly perturbed, meaning that the surface displacement $d$ is small compared to the reference height $h_0$, the capillary-driven thin film equation can be linearized\cite{Salez2012,Benzaquen2013} by letting $
H(\boldsymbol R,T)=1+{\mathcal Z}(\boldsymbol R,T)$, with $|{\mathcal Z}(\boldsymbol R,T)|\ll 1$,
where ${\mathcal Z}=d/h_0$  denotes the dimensionless surface displacement. This yields, at the lowest order in ${\mathcal Z}$, the dimensionless linear thin film equation:
\begin{eqnarray}
\left(\partial_T+\Delta^{2}\right) {\mathcal Z}(\boldsymbol R,T)&=&0\ ,
\label{LTFEad}
\end{eqnarray}
where $\Delta^{2}$ denotes the bilaplacian operator. Equation~\eqref{LTFEad} can be solved by deriving its Green's function $\mathcal G(\boldsymbol R,T)$. Proceeding as in a previous communication\cite{Benzaquen2013} yields  $ \mathcal G(\boldsymbol R,T)=\breve{\mathcal{G}}(\boldsymbol U,T)$ with for all $T>0$:
\begin{eqnarray}
\breve{\mathcal{G}}(\boldsymbol U,T)
&=& \frac 1{T^{(d-1)/4}}\ \phi(\boldsymbol U)\ ,  \label{Defphi}
\end{eqnarray}
where the function $\phi$ {depends only on} the dimensionality of the system $d\in\{2,3\}$:
\begin{eqnarray}
\label{auxi}
\phi(\boldsymbol U)&=&\frac{1}{(2\pi)^{(d-1)}}\,\int  \textrm{d}^{(d-1)}\boldsymbol Q\,{e^{-\left(\boldsymbol Q ^2\right)^2}e^{i\boldsymbol Q \cdot \boldsymbol U}}\ ,
\end{eqnarray}
and where we introduced the self-similar variable $\boldsymbol U= \boldsymbol R T^{-1/4}$. Note that the function $\phi$ can be written in terms of hypergeometric functions\cite{Benzaquen2013,Benzaquen2014,Backholm2014,Salez2012}. The solution to any summable initial perturbation $\mathcal Z(\boldsymbol R, 0)=\mathcal Z_0(\boldsymbol R)$ is simply given by the convolution $(\mathcal G*\mathcal Z_0)(\boldsymbol R,T)$. Assuming that the initial perturbation is \textit{rapidly decreasing} in space, namely for all $n\in \mathbb N$, $\lim_{|\boldsymbol R|\rightarrow \infty}|\boldsymbol R|^n\mathcal Z_0(\boldsymbol R)=0$, allows for writing  the solution as a series in which the different terms naturally decrease with time, and where remarkably: the higher the order, the faster the decrease.
Defining respectively the algebraic volume, $\mathcal M_0$, the first and second moments, $\boldsymbol {\mathcal M}_1$ and $ \mathfrak M_2$, of the initial perturbation $\mathcal Z_0(\boldsymbol R)$, as well as the Hessian matrix $\mathcal H_\phi (\boldsymbol U) $ of the function $\phi (\boldsymbol U)$, yields ${{\mathcal Z}}(\boldsymbol R,T)=\breve{{\mathcal Z}}(\boldsymbol U,T)$ with:
\begin{widetext}
\begin{eqnarray}
\breve{{\mathcal Z}}(\boldsymbol U,T)
&=&\frac{1}{T^{(d-1)/4}} \left[     \mathcal M_0 \,\phi(\boldsymbol U)     -\frac{ \boldsymbol {\mathcal M}_1\cdot \boldsymbol \nabla \phi(\boldsymbol U)  }{T^{1/4}}  +\frac1{2}  \frac{\mathfrak M_2 : \mathcal H_\phi (\boldsymbol U) }{ T^{1/2}}+ \text O \left(\frac1{T^{3/4}}\right) \right] \ , \label{3firstterms}
\end{eqnarray}
 where $\mathfrak A:\mathfrak B=\sum_{ij}\mathfrak A_{ij}\mathfrak B_{ij}$ is the tensor contraction of $\mathfrak A$ and $\mathfrak B$. In the 2D case, $d=2$, where the profile depends on only one spatial variable, Eq.~\eqref{3firstterms} simply becomes:
\begin{eqnarray}
\breve{{\mathcal Z}}( U,T)
&=&\frac{1}{T^{1/4}} \left[     \mathcal M_0 \,\phi^{\text{2D}}( U)     -\frac{  {\mathcal M}_1 \,{\phi^{\text{2D}}}'( U)  }{T^{1/4}}  +\frac1{2}  \frac{\mathcal M_2 \,{\phi^{\text{2D}}}'' ( U) }{ T^{1/2}}+ \text O \left(\frac1{T^{3/4}}\right) \right] \ . \label{Dev1Dmethods}
\end{eqnarray}
where $\mathcal M_n=\int\text d X' \,{X'}^n \mathcal Z_0(X')$. In the 3D case, the first moments of the initial perturbation $\mathcal Z_0(\boldsymbol R)$ read:
\begin{subeqnarray}
\slabel{Vol3D}\mathcal M_0&=& \int \text d^2 \boldsymbol R \, \mathcal Z_0(\boldsymbol R)\\
\boldsymbol {\mathcal M}_1&=& \int \text d^2 \boldsymbol R \,\boldsymbol R\, \mathcal Z_0(\boldsymbol R)\\
{\mathfrak M}_2&=&  \begin{pmatrix}
 \displaystyle \,\int \text dX\,\text dY \, X^2 \mathcal Z_0(X,Y)  &\displaystyle \,\int \text dX\,\text dY \, XY \mathcal Z_0(X,Y)   \smallskip\\
\displaystyle \,\int \text dX\,\text dY \, XY \mathcal Z_0(X,Y)  &\displaystyle \int \text dX\,\text dY \, Y^2 \mathcal Z_0(X,Y)
\end{pmatrix}  \ . \label{MatrixPHI2D}
\end{subeqnarray}
Letting the polar change of variables $\boldsymbol U=U_R(\cos\psi,\sin\psi)$, together with Eqs.~\eqref{MatrixPHI2D} yields:
\begin{subeqnarray}
\boldsymbol {\mathcal M}_1\cdot \boldsymbol \nabla \phi^{\text{3D}}&=& \int \text d R' \text d\alpha \,R'^2\cos(\alpha-\psi)\,  \mathcal Z_0( R',\alpha) {\phi^{\text{3D}}}'(U_R) \\
\mathfrak M_2 : \mathcal H_{\phi^{\text{2D}}} &=&\int \text d R' \text d\alpha \,R'^3 \mathcal Z_0( R',\alpha)\left[\cos^2(\alpha-\psi){\phi^{\text{3D}}}''(U_R)+\sin^2(\alpha-\psi)\frac{{\phi^{\text{3D}}}'(U_R)}{U_R} \right]\ . \hspace{1cm}
\end{subeqnarray}
Considering the averaged profiles over the angle $\psi$ yields $\langle   \boldsymbol {\mathcal M}_1\cdot \boldsymbol \nabla \phi^{\text{3D}}    \rangle_\psi=0$ and:
\begin{eqnarray}
\langle\mathfrak M_2 : \mathcal H_\phi^{\text{3D}} \rangle_\psi&=&\frac12\int \text d R' \text  \,R'^3 \mathcal Z_0( R')\left[{\phi^{\text{3D}}}''(U_R)+\frac{{\phi^{\text{3D}}}'(U_R)}{U_R} \right] \ .   \label{averagedPsiM2}
\end{eqnarray}
Note as well that, if the initial profile is axisymmetric, namely $\mathcal Z_0(R,\alpha)=\mathcal Z_0(R)$ then one has $\boldsymbol {\mathcal M}_1\cdot \boldsymbol \nabla \phi^{\text{3D}}=0$, and $\mathfrak M_2 : \mathcal H_{\phi^{\text{3D}}} =\langle\mathfrak M_2 : \mathcal H_\phi^{\text{2D}} \rangle_\psi$ as given by Eq.~\eqref{averagedPsiM2}.
In the manuscript, Eq.~\eqref{2D_expansion} is none other than Eq.~\eqref{Dev1Dmethods} where we have let $F_n(u)=\phi^{\text{2D}(n)}(u)$, and Eq.~\eqref{3D_expansion} is the angular average of Eq.~\eqref{3firstterms}, where $G_0(v)=\phi^{\text{3D}}(v)$, $G_2(v)={\phi^{\text{3D}}}''(v)+{\phi^{\text{3D}}}'(v)/v$, $\mathcal N_0=\mathcal M_0$ as given by Eq.~\eqref{Vol3D}, and $\mathcal
N_2=\int \text d R' \text  \,R'^3 \mathcal Z_0( R')$, consistent with Eq.~\eqref{averagedPsiM2}.\medskip

\vspace{1cm}

\end{widetext}

As mentioned in  the article, the convergence time can be computed analytically using the scheme proposed {previously}~\cite{Benzaquen2014} for the particular case of non-{zero} volume surface perturbations. Briefly, denoting by $\mathcal Z_\infty(\boldsymbol R,T)$ the dimensionless surface displacement in the long-term asymptotic regime, the dimensionless convergence time is determined by its intersection with the initial amplitude. Choosing the maximum amplitude over space as a reference value yields: $\max_{X}|\mathcal Z_\infty(X,T_{\text c})|=|\mathcal Z_{0}|_{\max}$. For the three-dimensional case, the method is naturally applied to the angularly averaged profiles. Using Eqs.~\eqref{3firstterms}, \eqref{Dev1Dmethods} and \eqref{averagedPsiM2} in the three particular cases relevant to the experiments presented in this article, one obtains:

\begin{subeqnarray} \label{TcTheoretical}
&&\text{2D asymmetric: \,\,\,\, } T_{\textrm{c}}\,\,\,\,=\,\,\,\,\left(\frac{\mathcal M_1 {\phi^{\text{2D}}}'_{\max}}{  \mathcal Z_{0\max}} \right)^2\\
&&\text{2D symmetric: \,\,\,\, }  \hspace{0.2cm} T_{\textrm{c}}\,\,\,\,=\,\,\,\, \left|\,\frac{\mathcal M_2 {\phi^{\text{2D}}}''(0)}{2\mathcal Z_0(0)}\,\right|^{\,4/3} \\
&&\text{3D: \,\,\,\, } \hspace{1.985cm}T_{\textrm{c}}\,\,\,\,=\,\,\,\, \left|\,\frac{\mathcal N_2 }{16 \pi\mathcal Z_0(0)} \, \right| \ .
\end{subeqnarray}

\bibliography{higherordercapillarylevelling}
\end{document}